\documentclass[preprint,showpacs,preprintnumbers,amsmath,amssymb]{revtex4}

\usepackage{graphicx}

\usepackage{color}

\usepackage{dcolumn}
\usepackage{bm}
\usepackage{float}

\newcommand{\be}{\begin{equation}}
\newcommand{\ee}{\end{equation}}
\newcommand{\bea}{\begin{eqnarray}}
\newcommand{\eea}{\end{eqnarray}}

\newcommand{\kb}{k_{\mathrm{B}}} 

\newcommand{\svlu}{\langle \sigma_{\mathrm{loss}}(U) \ v\rangle} 
\newcommand{\vmot}{V_{\mathrm{MOT}}} 
\newcommand{\vmt}{V_{\mathrm{MT}}} 
\newcommand{\nmot}{N_{\mathrm{MOT}}} 
\newcommand{\nmt}{N_{\mathrm{MT}}} 

\newcommand{\svl}{\langle \sigma_{\mathrm{loss}} v\rangle} 
\newcommand{\svt}{\langle \sigma_{\mathrm{tot}} v \rangle} 
\newcommand{\vp}{{\bar{v}}}
\newcommand{\svtn}{\left<\sigma_{\rm{tot}}^{(n)} \ {v}\right>}

\newcommand{\gloss}{\Gamma_{\mathrm{loss}}}
\newcommand{\gtot}{\Gamma_{\mathrm{tot}}}

\newcommand{\svloss}{\langle \sigma_{\mathrm{loss}} v \rangle}
\newcommand{\svtot}{\langle \sigma_{\mathrm{tot}} v \rangle}
\newcommand{\svtotexp}{\langle \sigma_{\mathrm{tot}} v \rangle_{\mathrm{exp}}}

\newcommand{\ig}{i_{\mathrm{g}}}
\newcommand{\igexp}{i_{\mathrm{g,exp}}}
\newcommand{\igNIST}{i_{\mathrm{g,NIST}}}

\newcommand{\nb}{n_{\mathrm{test}}}

\newcommand{\pb}{P_{\mathrm{test}}}
\newcommand{\px}{P_{\mathrm{x}}}

\newcommand{\pbs}{P_{\mathrm{base}}}
\newcommand{\parg}{P_{\mathrm{Ar}}}
\newcommand{\pxe}{P_{\mathrm{Xe}}}

\newcommand{\ma}{m_{\mathrm{t}}}
\newcommand{\mb}{m_{\mathrm{bg}}}

\newcommand{\tmin}{\theta_{\mathrm{min}}}

\newcommand{\ut}{U}
\newcommand{\utmax}{U_{\mathrm{max}}}
\newcommand{\udiff}{U_{\mathrm{d}}}
\newcommand{\uud}{\frac{U_{\phantom{d}}}{U_{\mathrm{d}}}}

\newcommand{\sbar}{\bar{\sigma}}
\newcommand{\vbar}{\bar{v}}
\newcommand{\mt}{m_{\mathrm{t}}}

\newcommand{\funiv}{p_\mathrm{QDU6}}

\begin{document}
\preprint{APS/??}
\title{Universality of Quantum Diffractive Collisions and the Quantum Pressure Standard}
\author{James L.~Booth$^{1,*}$, Pinrui Shen$^{2,*}$, Roman V.~Krems$^{3}$ and Kirk W.~Madison$^{2}$}
\affiliation{$^1$ Department of Physics, British Columbia Institute of Technology,\\
3700 Willingdon Avenue, Burnaby, B.C. V5G 3H2, Canada}
\affiliation{$^2$Department of Physics and Astronomy, University of British Columbia,\\
6224 Agricultural Road, Vancouver, BC, V6T 1Z1, Canada}
\affiliation{$^3$Department of Chemistry, University of British Columbia,\\
6224 Agricultural Road, Vancouver, BC, V6T 1Z1, Canada}
\affiliation{$^*$These authors contributed equally to this work.}

\date{\today}

\begin{abstract}
This work demonstrates that quantum diffractive collisions are governed by a universal law { characterized by a single parameter that can be determined experimentally.}  Specifically, { we determine a quantitative form of the  universal, cumulative energy distribution transferred to initially stationary sensor particles by quantum diffractive collisions.}  The characteristic energy scale corresponds to the localization length associated with the collision-induced quantum measurement, and the shape of the universal function is determined only by the analytic form of the interaction potential at long range.  Using cold $^{87}$Rb sensor atoms confined in a magnetic trap, we observe experimentally $\funiv$, the universal function specific to van der Waals collisions, and use it to realize a \emph{self-defining} particle pressure sensor that can be used for any ambient gas.  This provides the first primary and quantum definition of the Pascal, applicable to any species and therefore represents a fundamental advance for vacuum and pressure metrology.  The quantum pressure standard realized here is compared with a state-of-the-art orifice flow standard transferred by an ionization gauge calibrated for N$_2$. The pressure measurements agree at the 0.5\% level. 
\end{abstract}

\maketitle

\section{Introduction}

{There are many applications in physics that use the outcome of collisions of microscopic particles (sub-atomic particles, atoms or molecules) as a parameter to model complex physical behavior.  However, microscopic collisions are generally system-specific and depend on the collision details such as the total angular momentum, the quantum states of the collision partners and the collision energy. In certain limits, it is possible to obtain universal functions that describe the physics of microscopic collisions by a few parameters. For example, the cross section for insertion chemical reactions of molecules at ultra-cold temperatures is a universal function of the scattering length \cite{PhysRevLett.105.263203,doi:10.1103,doi:10.1021}. Finding such universal functions for different regimes of collision physics is necessary for applications bridging microscopic collisions with macroscopic phenomena.

In this work, we report a form of collision universality that occurs at ambient temperatures.  We demonstrate, both theoretically and experimentally, that the low energy behaviour of the cumulative energy distribution imparted to an initially stationary sensor particle embedded in a gas at thermal equilibrium is described by a universal function 
that depends only on (i) the analytic form of the interaction potential at long-range and (ii) the quantum diffraction energy, $\udiff \equiv \frac{4 \pi \hbar^2}{\mt \sbar}$ \cite{PhysRevA.60.R29}.  Here $\mt$ is the mass of the sensor particle and $\sbar$ is the thermally-averaged total collision cross section, including contributions from both elastic and inelastic scattering. We further demonstrate that $\sbar$ is independent of the short range interaction between the colliding particles with van der Waals long-range interactions. 

It is well known that collisions resulting in small momentum transfer are dominated by quantum diffractive scattering \cite{PhysRevA.60.R29,PhysRevA.80.022712}. Such collisions occur with small scattering angles $\theta \rightarrow 0$ and are predominantly determined by the long-range part of the interaction potential (see, for example, the discussion in Ref. \cite{Child}).  They are expected to be independent of the short-range interactions between the colliding particles. A qualitative relationship between the long-range interaction parameters and the scattering amplitude (i.e.~differential cross section) can be established by an analysis based on the Born approximation \cite{Child}; however, the resulting predictions cannot be used for the quantitative characterization of quantum diffractive scattering observables.  In the present work we analyze, by numerical calculations and experimental measurements, the dependence of quantum diffractive collisions on the long-range interactions and find a universal function for the cumulative energy distribution imparted to an initially stationary sensor particle, $\funiv$. This function is parametrized by a single, physical parameter, $U_d$, and through it by $\sbar$.  We show that, given the universal function, $U_d$ and, consequently, $\sbar$, can be extracted from the measurement of the  energy dependence of the thermally averaged, collisional energy exchange, without any input from theoretical calculations.  This is important because it is generally difficult, or even impossible, to compute $\sbar$ for complex molecular species with high precision using quantum scattering theory.}

Quantum diffraction universality (QDU) can be observed using a momentum or energy spectrometer provided one of the collision partners (the sensor particle) has a well-known initial energy distribution allowing the detection of the exceedingly low energy transferred.  
In this work, we observe QDU using laser-cooled $^{87}$Rb sensor atoms that are prepared in a magnetic trap with an average energy ($< 0.5$~mK) well below the characteristic quantum diffractive energy ($U_d \sim 10$~mK) for van der Waals interactions.  By measuring the trap loss rate as a function of trap depth, determined by the cumulative energy distribution after a collision, we observe the universal law specific to van der Waals interactions for a variety of atomic and molecular collision partners at room temperature.  These measurements provide a direct probe of the thermally averaged, total cross-section.

Using the universal function for van der Waals interactions, we demonstrate a self-defining flux sensor that provides a measurement-based, quantum mechanical definition of the Pascal applicable to any atomic or molecular species.  Specifically, the total cross section and density of the impinging particles can be found from a measurement of trap loss as a function of trap depth, described by $\funiv$.  This new pressure standard and the underlying theory of QDU is validated by measurements with N$_2$.  Following the conventional pressure standards comparison methodology, pressure readings from the quantum standard were compared to and found to agree with a those of a NIST-calibrated ionization gauge (IG) to within 0.5\% \footnote{The IG was calibrated by an orifice flow pressure standard which will be returned to NIST so that the calibration can be checked to complete this standards comparison. This is the conventional comparison protocol.}.

\section{QDU-based Metrology}

Vacuum measurement plays a central role in a wide range of scientific and industrial applications including residual gas analysis, semi-conductor device manufacture, and atmospheric modeling.  Until now, there has been no primary pressure standard for the high (HV) and ultra-high vacuum (UHV) regime ($< 10^{-7}$~Pa). Moreover, ionization-based gauges, used ubiquitously for measurements in this range, are plagued by well-known limitations. These include (i) the need to calibrate each gauge for each gas of interest, (ii) the loss of calibration due to device aging and exposure to gases \cite{Fedchak}, and (iii) their limitation to non-reactive species to avoid device contamination and subsequent measurement error.  Calibration loss is especially severe for residual gas analyzers (another ionization-based detector) rendering non-\emph{in situ} calibration efforts futile \footnote{J. A.~Fedchak (private communication, 2012).}.  The reliance on ionization gauges for metrology is equivalent to measuring distances with a meter stick whose length changes with use in an unpredictable way and whose scale is unknown for most objects being measured.  In addition, current state-of-the-art pressure standards are not primary, challenging to operate in the UHV, and fundamentally limited to measurements of inert gas only.

QDU eliminates all of these limitations by enabling the creation of a self-calibrating atomic sensor immune to sensor degradation and applicable to any species, overcoming a long standing and fundamental limitation of existing secondary pressure standards and of ionization based gauges.  Because it provides a true value for the measurement of particle flux or density and is based on immutable and fundamental atomic and molecular constants, the QDU sensor fits within the Quantum SI standards paradigm \cite{Goebel2005,Nawrocki2015}.  It provides a quantum definition of the Pascal that can be connected to all other pressure regimes using transfer standards such as spinning rotor gauges and ionization gauges.

The demonstration of QDU in the present work is a key development for atom-based pressure sensors. To date, some steps to realize our original proposal \cite{Patent,Madison2012} have been made. For example, magneto-optical traps (MOTs) have been used to perform vacuum measurements \cite{PhysRevA.80.022712,PhysRevA.84.022708,PhysRevA.85.033420,Yuan:13,Rowan2015}. { However, the accuracy of these measurements remains an open question because of complications including the non-negligible intra-trap two body collisions in a MOT, the large and unknown depths of the MOTs used, and the presence of both ground and excited electronic state atoms in the trapped ensembles.  Losses from optical dipole traps (ODT) have also been used to determine the density of a gas introduced into a vacuum \cite{Makhalov2016,Makhalov2017}.  In that work, estimates of the collision cross sections were made using the Landau-Lifshitz-Schiff approximation and pressure determinations of N$_2$ were within a factor of 2 of those made by an ionization gauge.  More recently, researchers at NIST have estimated that they will achieve absolute pressure measurements with an uncertainty of 5$\%$ using a combination of trap loss measurements and \textit{ab initio} calculations of cross-sections for the Li + H$_2$ system.  They proposed to extend this primary SI traceability to other species using a dynamics gas expansion system \cite{Jousten2017,Julia2017,Julia2018}. While systematic, this approach is limited to measurements of inert gases only.  Most notably, all prior work has relied on unverified estimates of the collision cross sections because the required data or theory of the complete interaction potentials is unavailable.} QDU eliminates any reliance on previous measurements of collision parameters or theory for the interaction potentials by providing an empirical measure of the total cross section. In this work we show that it provides a pressure determination at the level of 1\% and is applicable for any atomic or molecular species.

\section{Theoretical Predictions}

In this work, we use the trap loss rate of cold atoms induced by exposure to a room-temperature gas to observe QDU.  The trapped sensor particle collision rate is $\gtot = n \svtot $, where $n$ is the density of impinging test gas particles and $\svtot$ is the total collision rate coefficient. The brackets indicate an average over the Maxwell-Boltzmann (MB) speed distribution of the test gas particles.  Not every collision will induce loss from a trap of finite depth $\ut$ \cite{PhysRevA.80.022712,PhysRevA.84.022708}, and we define $\funiv$ as the probability that the sensor atom remains in the trap after the collision.  As $\ut \rightarrow 0$, $\funiv \rightarrow 0$ and the loss rate approaches the total collision rate.  Thus, we expand the loss rate for small $U$ in powers of the scaled trap depth
\bea
\gloss & = & n \svloss = n \svtot \cdot (1-\funiv),
\label{eq:universal-gloss}
\eea
with
\bea
\funiv & \equiv & \sum_{j=1}^{\infty} \beta_j  \left(\frac{U}{\udiff} \right)^j,
\label{eq:universal-curve}
\eea
where $\svloss$ is the velocity averaged collision loss rate coefficient, and $\udiff$ is the characteristic quantum diffraction energy. 
One of the key results of this work is the theoretical and experimental demonstration that $\funiv$ is a universal function with coefficients, $\beta_j$, that are independent of the short range details of the potential, independent of the strength of the van der Waals interaction (i.e.~the value for $C_6$), and independent of the masses of the trapped and incident particles. Because $\funiv$ is expected to be a universal function and $U_d$ is the only system-dependent parameter, consequently, $U_d$ must also be independent of the short-range interactions between the collision partners.

\begin{figure}[ht!]
\centering
\includegraphics[width=6.0in]{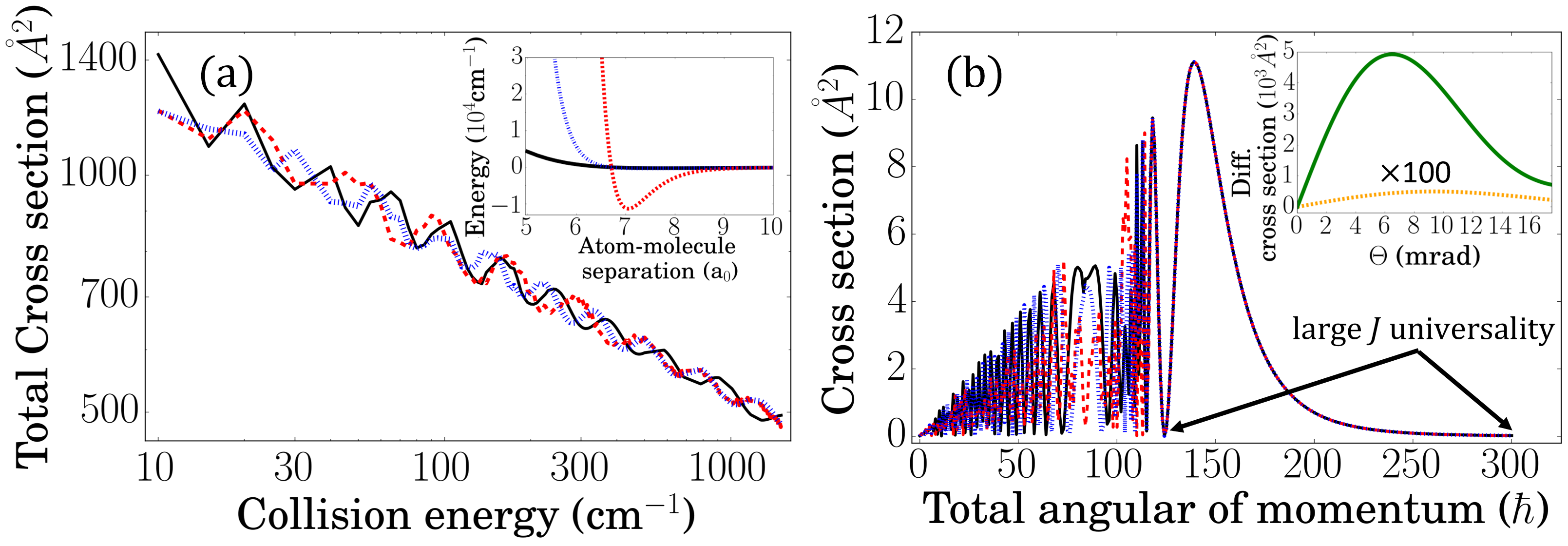}
\caption{Theoretical demonstration of collision universality.  Atom-molecule cross-sections vs.~collision energy in (a) are for three PESs (see inset) with different cores but with the same long range interaction.  {  The atom - molecule PESs illustrated in the inset have an anisotropic
long-range interaction with $C_{60}=350.24$ a.u. and $C_{62}=191.04$ a.u.
As described in Appendix B, at short range, these PES are represented by a
Legendre expansion with 7 terms. Each of these terms is modulated by a
distance-dependent factor, given by Eq. (B.8), leading to a family of
three drastically different PES.} The thermally averaged total cross section is the same (to within 0.6\%, see text) for all three PESs, despite the radical differences in the short-range interactions.  The cross-sections versus $J$ are shown in (b) for a collision energy of 300~cm$^{-1}$ and exhibit a universal shape above $J=125 \hbar$ and core-dependent oscillations below. The inset of panel (b) shows the cross sections for elastic (solid curve) and inelastic (dashed curve) scattering for $\theta < 10$ mrad. 
}
\label{fig:justification}
\end{figure}

\subsection{Relation between quantum diffraction energy and collision cross section}

\label{heisenberg}
{
In previous work considering predominantly elastic scattering \cite{PhysRevA.60.R29}, the characteristic quantum diffraction energy $U_d$ was defined as $\udiff \equiv \frac{4 \pi \hbar^2}{\mt \sigma}$, where $m_t$ is the mass of the trapped atom and $\sigma$ is the collision cross section.  The relationship between the collision cross section and $U_d$ is dictated by the uncertainty principle and is a consequence of the collision-induced sensor particle localization: a collision between two particles necessarily localizes the partners in real space to $\sbar$.  
Complementarity requires that this position localization of the sensor particle, $\Delta x \sim \sqrt{\sbar}$, results in a momentum distribution of characteristic minimum width $\Delta p \sim \hbar /\sqrt{\sbar}$.  
Therefore, the observation of the distribution of momenta imparted to the sensor particles (which we achieve by observing the trap loss probability versus trap depth revealing the sensor particle cumulative energy distribution post-collision) provides a direct probe of $\sbar$. Because particles are expelled from the trap by both elastic and inelastic collisions that impart energy exceeding the trap depth, $\sbar$ is the total collision cross section including the elastic and inelastic scattering contributions. For collisions induced by a gas at thermal equilibrium, the quantum diffractive energy, $U_d$ takes the form, 
\begin{eqnarray}
\udiff \equiv \frac{4 \pi \hbar^2}{\mt \sbar}, 
\label{Ud}
\end{eqnarray}
where $\sbar = \langle \sigma_{\rm tot} \rangle$ is a thermally-averaged total collision cross section. 

In the present work, we observe the trap loss rate due to a thermal ensemble of particles impinging on the trapped atoms. Thus,  the quantity of interest is the thermally-averaged collision rate coefficient, $\svtot$. 
Therefore, we define $\sbar$ in Eq. (\ref{Ud}) as $\sbar = \svtot/\vbar$, where $\vbar = \sqrt{2 \kb T / \mb}$ is the most probable relative velocity given the MB speed distribution of the incident particles at temperature $T$.
This choice for $\sbar$ is supported by our analytical model (Appendix D, Eq.~\ref{eq:svl_univ}) which shows that the loss rate coefficient, $\svl$, can be expressed as a polynomial in powers of the quantity $\left(m_t\cdot \svtot \ U\right)/ (4 \pi \hbar^2 \bar{v}) \equiv U/\left[4 \pi \hbar^2/ (m_t \bar{\sigma})\right]$.}
The precise shape of the energy distribution (i.e.~the values of the coefficients $\beta_i$ in Eq.~(\ref{eq:universal-curve})) scaled by this characteristic width depends on the long-range potential shape, and different analytic forms (e.g.~$C_n/r^n$ with $n>3$) lead to distinct universal functions constituting different universality classes.

\subsection{Numerical results}

In prior work on pressure broadening spectroscopy, evidence was found that the coherence kernel width, due to quantum diffractive collisions, is independent of the ratio of the perturber to active-atom mass and depends on the active-atom mass and elastic collision cross section \cite{PhysRevA.25.2550,PhysRevLett.50.331}.  Our work explores this further by (i) demonstrating the universality of both the thermally-averaged, total and loss cross sections and (ii) by providing the universal function that links the two.  To illustrate why the total cross section is universal, consider Fig.~\ref{fig:justification}(a) which shows the total cross section as a function of collision energy.  The three curves were computed using the time-independent coupled channel (CC) approach (described in Appendix B and in \cite{ArthursDalgarno1960,RomanBook}) for an atom-diatomic molecule collision for three different potential energy surfaces (PES).  Each PES, shown at a $90^{\circ}$ Jacobi angle of approach in the inset, has the same long range van der Waals potential but radically different short range core potentials, differing in depth by more than a factor of $10^4$.  The cross sections exhibit core-dependent oscillations super-imposed on a trend defined by the long-range part of the potential \footnote{For a long range potential varying as $C_n/r^n$ the trend is approximately a power-law, $\sigma(v) \sim \left( \frac{C_n}{\hbar v}\right)^{\frac{n-4}{5}}$, that only depends on $C_n$ (see Ref.~\cite{Child}).}.  The oscillations arise from the velocity dependent glory phase shift and therefore the locations of the maxima and minima are dependent on the short range physics \cite{Child}.  The effect of thermal averaging is clear: While an accurate prediction of the collision rate for a given velocity requires detailed knowledge of the core potential, averaging the cross section over one or more oscillations removes the core-dependent effects.  In particular, for the 3 different PESs, we find $\svtot = [0.361,0.361,0.363]\times 10^{-8}$~cm$^3$/s for the dark solid, dotted, and dashed PESs respectively.  Because all three PESs have identical long range character the thermally averaged total cross-sections are identical (differing by much less than 1\%) although the short range physics of the interactions and the corresponding inelastic collision rates are radically different.

The shape of $\funiv$ and corresponding loss cross section for small $U$ is independent of the short range part of the potential because inelastic and small impact parameter elastic collisions that probe the core always lead to large energy transfer and loss for shallow traps.  {Thus, the loss rate departs from the total collision rate due only to quantum diffractive collisions.  This departure is a direct measure of the low-angle scattering cross section which is expected to be independent of the short range potential \cite{Child}.}  Fig.~\ref{fig:justification}(b) shows the cross-section versus total angular momentum, $J$, for the three PESs at a collision energy of 300~cm$^{-1}$.  The curves exhibit the same universal shape, independent of the core potential above $J=125 \hbar$.  The scattering angles of such collisions are tiny ($<1$ mrad for $U=1$~mK and a collision energy of 300~cm$^{-1}$ \footnote{Collisions below $\tmin = \arccos \left(1- \frac{\mt \ut}{\mu^2 v^2} \right)$ (where $\mt$ is the sensor atom mass and $\mu$ is the reduced mass) do not lead to trap loss for an initially stationary atom (see Refs.~\cite{PhysRevA.80.022712,PhysRevA.84.022708}).}) where the differential cross section, shown in the inset of Fig.~\ref{fig:justification}(b), is dominated by large impact parameter elastic scattering, more than a 1000 times larger than inelastic scattering (arising primarily from low $J$ collisions) below 10~mrad.

\begin{figure}
\centering
\includegraphics[width=2.94in]{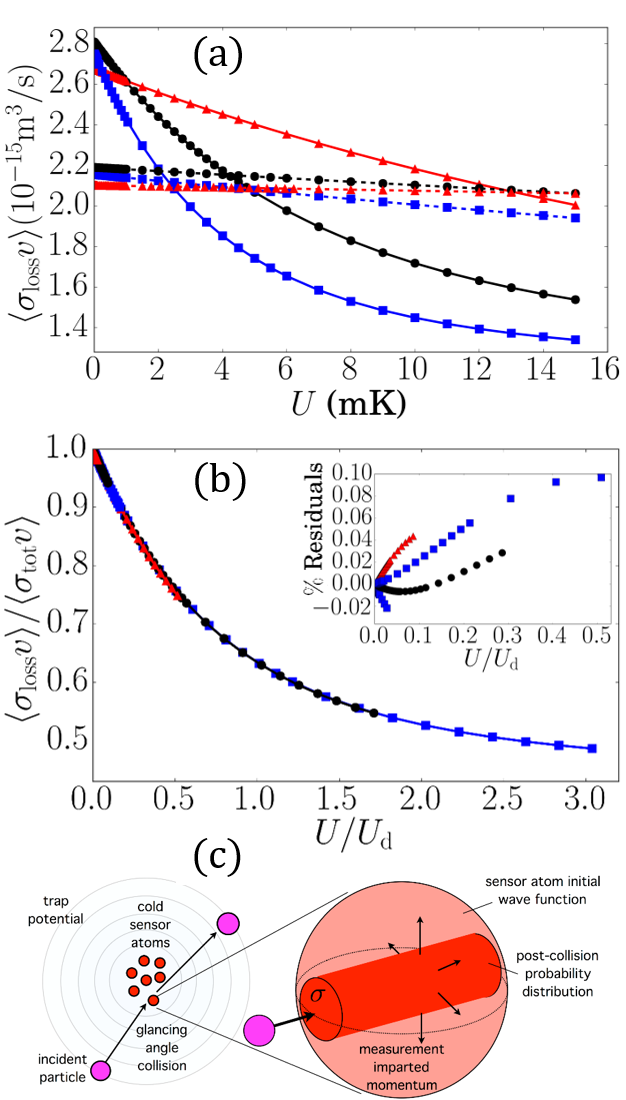}
\caption{Velocity averaged loss cross sections versus trap depth for He (red triangles), Ar (black circles), and Xe (blue squares) colliding with Li (dashed lines) and Rb (solid lines).  In (b) these loss rate coefficients are normalized by their value at $U=0$ and plotted versus the scaled trap depth.  All of the results collapse to the universal curve for $(1-\funiv)$ in Eq.~(\ref{eq:universal-curve}) with coefficients provided in Table~\ref{tab:betas}.  The inset shows the residuals for each calculation from the universal curve, and all are below $0.1 \%$ for trap depths up to $U = 2.2$~mK (the range of our measurements).  A schematic of the quantum measurement produced by a glancing collision is shown in (c).  Initially, the cold sensor atom has an large position uncertainty, and the collision localizes it to a small region of size $\sbar$.  This results in a broadened momentum distribution and a non-zero probability that the sensor particle escapes the trapping potential.  The shape of the loss rate curve in (b) is tied to the momentum distribution post collision.}
\label{fig:sc}
\end{figure}

Fig.~\ref{fig:sc}(a) shows quantum scattering calculations for the loss rate coefficient, $\svloss$, given a gas of He, Ar, and Xe at 21~C and trapped Rb or Li atoms. The interatomic interaction was modeled as a Lennard-Jones potential, $V(R) = 4 \epsilon \left[ \left( \frac{r_0}{R}\right)^{12} - \left( \frac{r_0}{R}\right)^{6} \right] = \frac{C_{12}}{R^{12}} - \frac{C_6}{R^6}$, where $\epsilon$ is the depth of the potential well and $r_0 = (C_{12}/C_6)^{1/6}$ is the range of the core repulsion.  The $C_6$ values are from Ref.~\cite{Andrei201042}, and the potential depth was {fixed at} $\epsilon = 50$~cm$^{-1}$.  The normalized loss rate coefficients, $\svloss / \svtot$, versus $U/\udiff$, are shown in Fig.~\ref{fig:sc}(b). All the results collapse to the universal curve, $(1-\funiv)$, with coefficients given in the first line of Table~\ref{tab:betas}.  The universal curve coefficients, $\beta_j$, are obtained by the best fit to these six calculations.  The residuals between the universal curve and the individual QS computations are shown in the inset and are all below $0.1\%$ for trap depths up to $U = 2.2$~mK (the range of our measurements).

The trap depths explored in this calculation were from 0 to 15~mK and are far beyond those realized experimentally (0.2 to 2.2~mK).  The corresponding scaled trap depth values ($U/\udiff$) differ depending on the total cross sections.  For example, the maximum scaled trap depth for Rb-He at 15 mK was 0.06 compared to the Rb-Xe value of 3.0.  The purpose of exploring the behavior of the universal curve at values of $U/\udiff \ge 1$ was to demonstrate the convergence of the series expansion insuring that it faithfully captures the universal behavior for our experimentally accessible values $U/\udiff \le 0.4$.

\begin{center}
\begin{table}[b!]
\begin{tabular}{|c|c|c|c|c|c|c|c|c|}
\hline
T (K) & $\epsilon$~(cm$^{-1}$) & $\beta_1$ & $\beta_2$ & $\beta_3$ & $\beta_4$ & $\beta_5$ & $\beta_6$ &
$\funiv^{(0.3)}$ \\
\hline
\hline
 294 & 50  & $0.6730(7)$ & -0.477 (3) & 0.228 (6) & -0.0703(42) & 0.0123 (14) & -0.0009 (2) & 0.165  \\ 
\hline
\hline
\hline
294 & 50  & $0.6754$ & -0.4992 (2) & 0.2775 (6) & -0.1165 (7) & 0.0321 (4) & -0.00413 (8) & 0.164  \\
273 & 50  & $0.6754$ & -0.4996 (2) & 0.2779 (6) & -0.1165 (7) & 0.0319 (4) & -0.00408 (8) & 0.164  \\
373 & 50  & $0.6749$ & -0.4970 (2) & 0.2759 (5) & -0.1165 (6) & 0.0326 (4) & -0.00433 (8) & 0.164 \\
40 & 50  & $0.6754$ & -0.4991 (7) & 0.2687 (14) & -0.1011 (12) & 0.0228 (4) & -0.00223 (6) & 0.164 \\
\hline
3 & 50 & $0.6471 (7) $ & -0.4317 (21) & 0.1889 (23) & -0.0516 (11) & 0.0078 (2) & -0.00048 (2) & 0.160 \\
988 & 50 & $0.7051 $ & -0.5389 (1) & 0.3086 (3) & -0.1369 (5) & 0.0421 (3) & -0.00640 (9) & 0.170 \\
\hline
294 & 500  & 0.6736 & -0.4976 (2) & 0.2765 (5) & -0.1161 (7) & 0.0320 (4) & -0.00411 (8) & 0.164  \\
294 & 5000  & $0.6736$ & -0.4977 (2) & 0.2763 (7) & -0.1157 (8) & 0.0318 (4) & -0.00408 (9) & 0.164 \\
294 & 50000 & $0.6736$ & -0.4977 (2) & 0.2767 (5) & -0.1162 (7) & 0.0320 (4) & -0.00412 (8) & 0.164 \\
\hline
\end{tabular}
\caption{The first six coefficients of $\funiv$, the universal curve (Eq.~(\ref{eq:universal-curve})) for van der Waals collisions at room temperature, are shown in the first line extracted from the best fit to the calculations in Fig.~\ref{fig:justification}(c).  $\beta_i$ values fit to calculations for just Rb with Ar at different temperatures and potential depths follow.  Based on the values of $\funiv$ at $U/\udiff=0.3$ (in the last column), the results are insensitive over a large temperature range (from 40 to 373~K) and to radical changes of the core potential depth (the last three lines).  However, at 3~K, the thermal average is too narrow and does not sample a large enough velocity range to average away the cross section oscillations (see Fig.~\ref{fig:justification}), and at 988~K, the temperature is too high and involves collisions at very large velocities whose cross section is influenced by the core shape of the potential.  In each case, both the coefficients and values for $\funiv$ are observed to deviate by more than 1\% from the room-temperature thermal average.  Despite this deviation, the systematic error in $\svtot$ and $\nb$ (the density of impinging particles) that would result by fitting trap loss induced by collision partners at 3K or at 988 K to the universal curve derived for $T=294$~K (the first row) is below 0.2\%.
}
\label{tab:betas}
\end{table}
\end{center}

{
Prior work, which is eloquently presented and summarized in Ref~\cite{Child}, has predicted that the scattering amplitude for elastic collisions at small angles can be written approximately in terms of the total elastic collision cross section. 
It is important to note, however, that using the results of Ref.~\cite{Child} to derive $\funiv$ yields a form similar to Eq.~(\ref{eq:universal-curve}), but with incorrect coefficients. We believe that the discrepancies arise, in part, because the velocity averaged cross section is influenced by partial wave mixing which is necessarily truncated by the commonly adopted approximations used for analytical modelling of small angle scattering (cf., Appendix D).  In this work, we use rigorous quantum scattering computations to obtain the correct coefficients and to explore their universality.}

To check the parameter range over which QDU applies for van der Waals collisions, the calculation for Rb sensor atoms and Ar gas was repeated at a variety of different temperatures and potential depths.  When the depth of the potential is varied by a factor of 1000, corresponding to a radical change of the short range physics, there is negligible variation of the universal coefficients and the trap loss rate at small depth, quantified by $\funiv(U/\udiff = 0.3)$ (see rows 8-11 of Table~\ref{tab:betas}).  The calculations for Rb with Ar gas temperatures in the range from 40~K to 373~K also show remarkably little variation (rows 2-5 of Table~\ref{tab:betas}).  

QDU breaks down in two important limits.  At very low temperatures, the MB distribution is so narrow that velocity averaging no longer eliminates the core dependent variations.  At high temperature, the high energy tail of the MB distribution of relative velocities can overlap the region above $v^* = 4 \epsilon r_0/\hbar$, where the trend for the velocity dependent cross section is significantly influenced by the core potential \cite{Child}.  In these two limits, the fitted coefficients and values for $\funiv$ at $U/\udiff=0.3$ deviate by more than 1\% from the room-temperature average.  Surprisingly, the systematic error in $\svtot$ and $\nb$ (the density of impinging particles) that would result by fitting the trap loss induced by collision partners at 3~K or at 988~K to the universal curve derived for T = 294~K is below 0.2\%.

\section{Experimental realization}
The experimental setup employs a test vacuum chamber housing an ionization gauge (IG) and the cold-atom sensor ensemble created by a 3D MOT, a standard six-beam magneto-optic trap (see Appendix A for details.). The 3D MOT is loaded by a flux of cold $^{87}$Rb atoms entering from a secondary 2D MOT chamber through a low conductance differential pumping tube, and ambient gas is introduced into the test section through a leak valve.
	
To observe QDU, the Rb atoms are transferred from the 3D MOT into a magnetic trap (MT).  The sensor atoms in the MT are in a single quantum state and the trap depth is set by radio frequency emission from an antenna (described in Appendix A). The trap loss rate due to collisions with the test species introduced through the leak valve is measured as a function of trap depth, $\gloss(U)$, Eq.~(\ref{eq:universal-curve}), at a fixed gas density. The baseline loss rate, associated with residual gases in the vacuum system and due to Majorana spin-flip losses \cite{Sukumar1997,Brink2006}, was measured as a function of trap depth and subtracted from the measurements so that $\gloss$ is attributed to the test gas alone. The density of the test gas was monitored using the (uncalibrated) IG readings, $\pb = \ig \nb \kb T$. Here, $\ig$ is the unknown gauge calibration factor (a species-dependent response) of the IG for the specific test gas, $k_{\rm{B}}$ is Boltzmann's constant, and $T$ is the temperature of the test gas. Provided the IG response is linear in the test gas density, $\nb$, variations in the density during the measurement process can be normalized away. Specifically,  we construct the quantities
\bea
\frac{\gloss(U)}{\pb /\left( \kb T \right)} & =& 
\frac{\svtot}{\ig}\left[1 - \sum_{j=1}^{\infty} \beta_j  \left(\frac{U}{\udiff} \right)^j \right]
\label{eq:gloss_expt}
\eea
that are fit to the QDU universal curve (on the right hand side) using two free parameters, $\svtot$ and $\ig$ ($\udiff$ is determined by $\svtot$ see the values in Table~\ref{tab:parameters}).  This construction has the advantage that the test gas pressure need only remain constant during a single MT lifetime measurement or only during a single MT hold time duration if the initial MT number is known.  Shot-to-shot pressure variations are normalized out by dividing $\gloss$ by $\pb$.

\section{Universal pressure standard}
\begin{center}
\begin{figure}[!ht]
\includegraphics[width=6.0in]{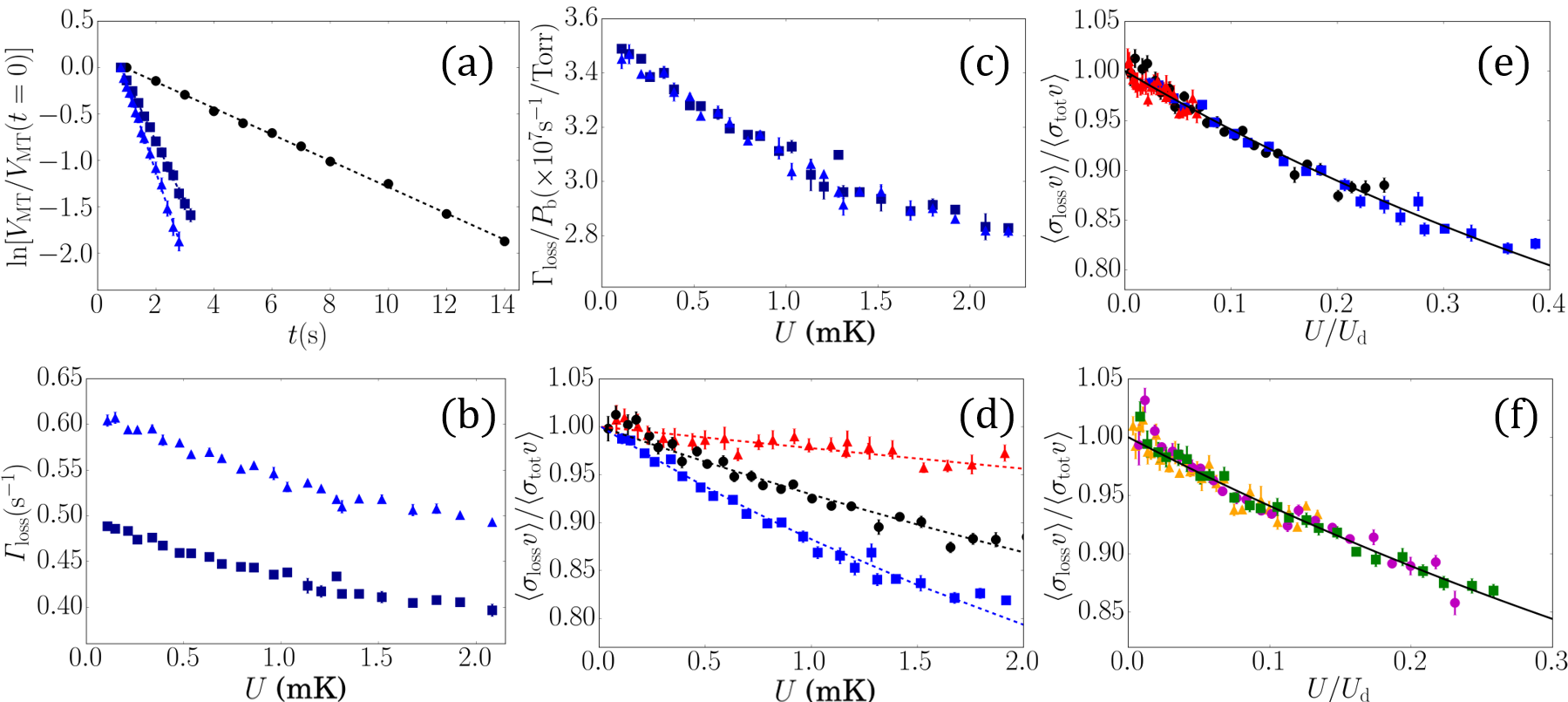}
\caption{(color online) Experimental results on trap loss rate universality.  In (a), the decay of the MT population with hold time is measured by recording the fluorescence upon MOT recapture of the $^{87}$Rb sensor atoms and normalizing it by the fluorescence after a negligibly short hold time.  In the absence of introduced gas, the ensemble exhibits an exponential decay (black circles) due to a variety of loss mechanisms including collisions with the residual background gases.  The decay slope steepens when an Ar partial pressure is added (square and diamonds).  In (b), the trap loss rate increase for $\pxe \sim 14.5$ (squares) and $18$ nTorr (triangles) are shown for different trap depths. These loss rates divided by the pressure reading for each measurement are shown in (c), and these values are then averaged and normalized by the extrapolated loss rate at zero trap depth and shown as blue squares in (d) along with similar results for He (red triangles) and Ar (black circles).  The data in (d) are shown in (e) with the abscissa scaled by $\udiff$ and fall on the universal curve (solid line).  The normalized loss rate versus scaled trap depth is also shown for collisions with molecules: H$_2$ (orange triangles),  N$_2$ (green squares), and CO$_2$(magenta circles) in (f).
}
\label{fig:experimentalresults}
\end{figure}
\end{center}
Figures~\ref{fig:experimentalresults}(a)-(f) show the experimental results. Panel (a) provides an example of the decay rate measurement (relative atom number in the MT as a function of the holding time for Rb-Ar collisions). The lowest (residual) decay rate corresponds to the loss rate measured with no Ar gas added and results from collisions with residual gases in the vacuum and from Majorana spin-flip losses from the MT.  Using a residual gas analyzer, we confirmed that as long as the residual gas composition is unchanged, this residual decay rate is constant, allowing it to be removed from the measurements to isolate $\gloss$. 
The other data in (a) show the atom loss from the MT for $\parg =$ 8.9 and 11 nTorr (1.2$\times 10^{-6}$~Pa and 1.5$\times 10^{-6}$~Pa). 

Panel (b) shows the MT loss rate due to Xe gas, extracted from curves like those shown in (a), as a function of trap depth for $\pxe =$ 14.5 and 18 nTorr (1.9$\times 10^{-6}$~Pa and 2.4$\times 10^{-6}$~Pa).
These same data normalized by pressure, $\gloss(U)/\pb$, shown in (c), verify the validity of Eq.~(\ref{eq:gloss_expt}).  After pressure normalization, the data are then scaled by the extrapolated loss rate at zero trap depth, $\gloss(0)$.  The loss rates, $\gloss(U)/\gloss(0) = \svloss/\svtot$, for Rb with He, Ar, and Xe are shown in (d).  Both the maximum loss rate, $\svtot$ and the shapes of the trap loss rate versus trap depth curves are different for each species.  Panels (e) and (f) show the plots of the normalized loss rate, $\svloss/\svtot$, as a function of the scaled trap depth $U/\udiff$ for atomic (He, Ar, and Xe) and molecular gases (H$_2$,  N$_2$, and CO$_2$), respectively.  The 
data sets all follow the
same QDU curve providing experimental verification of universality.  For each collision species, the values for $\udiff$ and $\ig$ were found by fitting the data to Eq.~(\ref{eq:gloss_expt}), and the results are shown in Tab.~\ref{tab:parameters}.
\begin{table}[t!]
\begin{tabular}{|c|c|c|c|c|}
\hline
&$\svtotexp (10^{-15}$m$^3/$s)  &{$\udiff$} (mK) &$\igexp$ & $\igNIST$ \\ \hline
He&2.40 ($\pm 5.0\%$)&32.3&0.163 ($\pm 4.9\%$) & - \\
Ar &2.77 ($\pm 1.8\%$)&8.9&1.238 ($\pm 2.1\%$) & -\\
Xe &2.71 ($\pm 1.4\%$)&5.0&2.511 ($\pm 1.5\%$) & -\\
$\text{H}_2$ &5.09 ($\pm 2.9\%$)&21.5&0.559 ($\pm
3.2\%$) & -\\
$\text{CO}_2$ &2.79 ($\pm 1.3\%$)&8.3&0.958
($\pm 1.6\%$) & -\\ 
$\text{N}_2$ &3.11 ($\pm 1.6\%$)&9.4&0.943 ($\pm 2.0\%$) & \textcolor{blue}{0.94 ($\pm 2.8\%$)}\\ \hline
\end{tabular}
\caption{Experimentally determined total cross sections ($\svtotexp$) and gauge calibration factors ($\igexp$) extracted from fitting trap loss data (Fig.~\ref{fig:experimentalresults})
to the universal curve (Eq.~(\ref{eq:gloss_expt})). The values of $\udiff$ are calculated based on $\svtotexp$.
The comparison between the $\igexp$ calibrated by the QDU pressure standard and by NIST's orifice flow standard shows excellent agreement. The uncertainties include both the fitting uncertainty and the uncertainty due to ensemble heating, and the precision of the QDU sensor is limited by the amount of data taken and the range of trap depths used (see text).
Only $\funiv$ and the values for $\svtotexp$ are needed for a $^{87}$Rb cold atom pressure standard.
}
\label{tab:parameters}
\end{table}

Current state-of-the-art pressure standards (known as orifice flow standards) operate for inert gases (e.g.~N$_2$), and inter-standard comparisons are carried out by calibrating an ionization gauge at one standard and then shipping it to a second standard for calibration comparison \cite{Fedchak}.  This same procedure was followed here to demonstrate the accuracy of the QDU primary quantum standard. An ionization gauge, calibrated by NIST for N$_2$ gas, was attached to our standard. 
As shown in Tab.~\ref{tab:parameters}, the NIST calibration factor of $\igNIST=0.94$ ($\pm 2.8$\%) is in excellent agreement with the value found with our quantum standard, $\igexp =0.943$ ($\pm\ 2\%$). The N$_2$ measurements were carried out after carefully preconditioning the IG \cite{Fedchak}.  That is, prior to calibrating the IG, the system was brought up to a pressure of $10^{-4}$~Torr of N$_2$ for one hour, then re-evacuated back to the system base pressure. This insures that N$_2$ gas saturates the IG filament so that only this species is emitted during the measurements. If this step is not performed then the measured gauge factor will vary over time, and the calibration comparison will be poor.   While the calibration of the $\ig$ for different species for the IG is a welcome outcome of the use of the QDU pressure standard, it is not central to the standard's operation.  
The values for $\svtot$ (provided in Tab.~\ref{tab:parameters}) and the universal coefficients $\beta_j$, alone, define the pressure standard.

\section{Conclusions}

{
In conclusion, we report the first measurement of the cumulative energy distribution imparted to an initially stationary sensor particle associated with quantum diffractive collisions. Based on quantum scattering calculations and experimental measurements
of the loss rate of trapped $^{87}$Rb atoms due to collisions with different
background gases, He, Ar, Xe, H$_2$, N$_2$, and CO$_2$, we determined the universal function describing quantum diffractive collisions by a single, experimentally measurable parameter. 
The universal function for the trap loss rate can be used to determine thermally averaged collision rates and the density of the ambient gas without input from other measurements or theoretical calculations.
The particular focus of this work is on determining and illustrating
the universality of $\funiv$ defined as the average cumulative energy distribution 
function of the sensor atoms after collisions with neutral atoms and molecules, characterized
by van der Walls interactions.} 

We use $\funiv$ to realize a self-defining pressure sensor and the first primary pressure standard for the high and ultra-high vacuum regimes applicable to any atomic or molecular species. This fundamental pressure definition can be connected to all other pressure regimes through the use of transfer standards, as is common practice for pressure metrology. Proof of universality and of the accuracy of this method is that the density extracted for N$_2$ is within 0.5\% of a measurement with a NIST calibrated ionization gauge (IG). 


Since QDU is a manifestation of the uncertainty principle and a consequence of the collision-induced sensor particle localization, it occurs for any interaction and applies to collisions of elementary particles, nuclei, atoms and molecules. Other long range interaction potentials (of the form $C_n/r^n$ with $n>3$) are characterized by similar but distinct universal functions constituting different universality classes for QDU (see Appendix D).  Future work will explore the universality for potentials with $n=4,5$, relevant for loss rates measurements of trapped molecules and ions from shallow traps \cite{Lambrecht2017,Schaetz2017,PhysRevLett.118.093201}.


\vskip 0.4in
We acknowledge financial support from the Natural Sciences and Engineering Research Council of Canada (NSERC / CRSNG) and the Canadian Foundation for Innovation (CFI). This work was done at the Center for Research on Ultra-Cold Systems (CRUCS).  P.S.~acknowledges support from the DFG within the GRK 2079/1 program.  We also wish to thank Takamasa Momose for supplying the Xe gas, D.A.~Steck for useful discussions, O.~Toader for a close reading of the manuscript, and James Fedchak of NIST for essential advice about IGs and for providing the calibrated IG.

\appendix
\section{Experimental details}

\begin{figure}
\begin{center}
\includegraphics[scale=1.0]{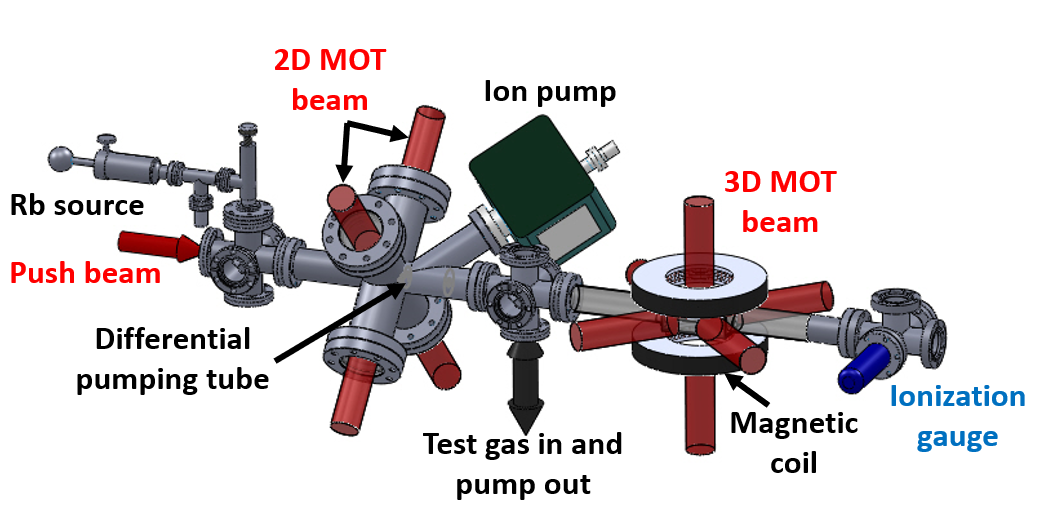}
\end{center}
\caption{A schematic of the experimental apparatus. Rb atoms are loaded from a vapor in the 2D MOT region separated from the 3D MOT section via two, low conductance, differential pumping tubes.  Atoms from a cold jet produced by the 2D MOT are captured in the 3D MOT, then transferred to the magnetic trap (MT). The test gas is introduced into the vacuum system through a leak valve, backed by a turbomolecular pump. The design of the vacuum system insures that the pressure of the test gas is stagnant in the 3D MOT section and there are no pressure gradients between the MT and the ionization gauges.}
\label{fig:Apparatus}
\end{figure}

The experimental setup, shown in FIG.~\ref{fig:Apparatus}, consists of a vacuum system composed of two sections: a 2D MOT region for collecting the sensor atoms, and a test section containing a 3D MOT, an ionization gauge (IG), and a turbo-molecular pump backed leak valve. 
The 3D MOT is loaded by a flux of cold $^{87}$Rb atoms entering from a secondary 2D MOT chamber through a low conductance differential pumping tube. This design separates the atom source in the 2D MOT section where the base pressure was $1\times10^{-7}$ Torr (set by the vapor pressure of Rb) from the 3D MOT test section with a base pressure of $2 \times 10^{-10}$ Torr.  Gas is introduced into the test section through the leak valve.  The vacuum system is carefully designed to insure that there is no pressure gradient between the IG and the trapped atom ensemble in the test section.

Two lasers, a pump and a repump, are used for cooling and trapping Rb in the 2D and 3D MOTs.  The pump laser is tuned 12 MHz below the D$_2$ ($^2$S$_{1/2}$ $\rightarrow$ $^2$P$_{3/2}$), $F=$~2$-$3$^{\prime}$ transition, while the repump laser is resonant with the $F=$~1$-$2$^{\prime}$ transition. The pump laser beam (2.2~cm in diameter) is split into three beams and retro-reflected to create the 3D MOT.  The repump light is incident on the 3D MOT volume via a single laser beam (2.2~cm in diameter). The pump and repump laser powers incident on the trapping region were 100 mW and 2 mW, respectively. The magnetic field configuration is a spherical quadrupole with an axial gradient of 13.6 G/cm.

The measurement cycle begins by loading $\nmot = 10^7$ atoms into the 3D MOT, as determined via the MOT fluorescence. The low atom number initially loaded into the MOT was selected to insure that the photodetector reading ($V_{\rm{MOT}}$) is linear in the atom number [37]. After loading, the atoms are cooled and transferred into the $F = 1$ state by changing the pump laser frequency tuning from 12 to 60 MHz below resonance, waiting 50~ms, and then extinguishing the repump light while leaving the pump light on for 4~ms.  The magnetic trap (MT) is established after the pump light is extinguished by increasing the axial magnetic field gradient to 272 G/cm, depending on the maximum trap depth desired. This procedure captures $\nmt = 2\times 10^6 \ (\pm 1.2\%)$ atoms in their $|F = 1, M_{\rm{F}}=-1\rangle$ state in the MT while ejecting atoms in the other $M_{\rm{F}}$ states. The magnetically trapped atoms are then held in complete darkness for a time interval, $t$, during which some of the atoms are lost due to collisions with particles in the vacuum. At the end of this ``holding'' time, the atoms are subjected to a RF field that sets the trap depth by ejecting all atoms in the MT whose energy is above  E $=$ $h\nu_{\rm{RF}}$ (Here $\nu_{\rm{RF}}$ is the frequency of the RF field.).  {The typical hold time is less than 10~s, while the 2-body collision rate for the cold sensor atoms in the magnetic trap is smaller than $10^{-5} \; \mathrm{s}^{-1}$.  Thus the sensor atoms behave completely independently for the duration of the measurement with an intra-trap thermalization timescale that is much longer than the hold time.}  The remaining atoms are recaptured in the 3D MOT and their fluorescence, $\vmt$, is recorded. This measurement is normalized by the 3D MOT fluorescence, $\vmot$, just before transfer to the MT. This ratiometric measurement helps to minimize the effect of shot-to-shot variations in the initial atom number loaded in the 3D MOT.  After a series of measurements with different holding times are carried out, the MT loss rate ($\gloss$) is extracted from,
\bea
\frac{\nmt}{\nmot} \propto \frac{\vmt}{\vmot}&=&\left.\frac{\vmt}{\vmot}\right|_{(t = 0)}e^{-\gloss t}.
\label{eq:vmtvstime}
\eea
Collision-induced heating of the sensor ensemble and an overestimate in the loss rates was avoided by limiting the hold times such that the fraction of heated atoms in the remaining ensemble ($f=1-e^{(\gtot - \gloss)t}$) was always less than 20\%.
Initially, the loss rate versus trap depth at the apparatus base pressure is recorded. These losses are due to common vacuum system species such as $\rm{H}_2$ and CO [35,36]. In addition to external gas collisions, the baseline loss rate includes Majorana losses and/or 2- and 3-body intra-trap losses, $\Gamma_0$. The trap-depth dependent baseline loss rate is,
\begin{eqnarray}
\Gamma_{\rm{base}}&=&\Gamma_0+\sum_{i=\rm{H}_2, \rm{CO}, ...} n_{i}\svlu_{\mathrm{Rb}-i},
\label{eq:Gamma_base}
\end{eqnarray}
The corresponding baseline ionization gauge (IG) pressure reading is, $\rm{P}_{\rm{base}}$. The measurements for $\Gamma_{\rm{base}}(U)$ are fit to a polynomial in trap depth (corresponding to a sum of trap-depth independent loss rates and a linear combination of universal laws with different scaled trap depths), constituting the baseline rate to be removed from the subsequent readings when the test gas species are introduced into the vacuum system. It was assumed that the residual gas composition that produced $\rm{P}_{\rm{base}}$ and the corresponding shape of $\Gamma_{\rm{base}}$ versus trap depth remained constant over the duration of the experiment.
Each test species (He, Ar, Xe, H$_2$, N$_2$, or CO$_2$) was introduced so that the IG pressure reading, $\px$, was at least 5 times higher than the base pressure. The subsequent measured loss rates, $\Gamma_{\rm{meas}}$, over a range of magnetic trap depths, $U$, was recorded.  
\begin{eqnarray}
\Gamma_{\rm{meas}}(U)&=& \Gamma_{\rm{base}}(U) + \Gamma_{\rm{loss}}(U) \nonumber \\
&=& \Gamma_{\rm{base}}(U) + n_{\rm{x}}\svlu_{\mathrm{Rb}-x}.
\end{eqnarray}
Subtracting the baseline loss rate from the measured loss rate yielded the loss rate caused by the test species, $\Gamma_{\rm{loss}}$, and the corresponding IG pressure reading attributed to the test gas was $\pb=\px-\pbs$. 

The test species pressure was then increased and the procedure was repeated, providing two data sets at two different test gas densities. When normalized by the test pressure, $\pb$, these two data sets overlapped (FIG.~3 (c) shows the Rb-Xe data), indicating the IG readings are linear in the test gas density. {Thus, we follow Eq.~(\ref{eq:gloss_expt}) to combine with the three experimentally measured quantities ($\gloss$, $T$, $\pb$) at a series of trap depths to get the fitting parameters, $\svtot$ and $\ig$. Comparing the best fit $\ig$ from the quantum pressure standard to the value reported from NIST, calibrated to their orifice flow standard, verifies the accuracy and precision of the form of the cumulative energy distribution due to QDC reported here.}

%

\subsection{Magnetic trap depth control}

$^{87}$Rb Atoms were loaded into the magnetic trap in the $|F = 1\  M_{\rm{F}}= -1\rangle$ state relative to the local field. These atoms evolved freely in the trap over the hold time until being recaptured and imaged in a MOT. We assume that the atoms with energy, $E$, travel out to a spatial location where their kinetic energy is zero and their potential energy is,
\begin{eqnarray}
E&=& -\vec{\mu}\cdot \vec{B} + mgz=h \mu_Bg_F|M_F|\left(\frac{dB}{dz}\right)\sqrt{\frac{x^2}{4}+\frac{y^2}{4}+z^2}+mgz.
\label{potential}
\end{eqnarray}
Here, the first term is the magnetic potential energy, and the second is the gravitational potential energy of the atom. The magnetic trap coils are arranged so that the axial B-field gradient, $\left(\frac{dB}{dz}\right)$, is aligned along the vertical- or z-direction. For a spherical quadrupole field, the axial gradient is twice the radial gradient, and the field is zero at the center of the two coils, $\vec{r} = 0$. In this coordinate system the gravitational potential energy is taken as zero at $z = 0$. 

For convenience, the axial gradient can be expressed as,
\begin{eqnarray}
\left(\frac{dB}{dz}\right) &=& b^{\prime} I
\end{eqnarray}
where $I$ is the current in the trapping coils. There is a minimum current required, $I_0$, to support the weight of the atoms against gravity,
\begin{eqnarray}
I_0 &=& \frac{mg}{h \mu_B g_F |M_F| b^{\prime}}.
\end{eqnarray}
The depth of the magnetic trap confining the Rb atoms was set by a radio-frequency (RF) B-field created by a single loop coil placed below the trapping region. The driving signal to the loop was frequency modulated over the range $[\nu_{\rm{min}}, \nu_{\rm{max}}]$ for the last 700 ms of each hold duration in the magnetic trap. For each RF frequency, $\nu$, there is a corresponding oblate spheroid surface where the RF field is resonant with $|F\ M_F\rangle \rightarrow |F\ M_F\pm1 \rangle$ magnetic dipole atomic transition. 
\begin{eqnarray}
h\nu &=& h\mu_B g_F \left(\frac{dB}{dz}\right) \sqrt{\frac{x^2}{4}+\frac{y^2}{4}+z^2} \nonumber \\
&=& \frac{mg} {|M_F|} \frac{I}{I_0}\sqrt{\frac{x^2}{4}+\frac{y^2}{4}+z^2}
\end{eqnarray}
Atoms with sufficient energy to traverse this surface will, with high probability, make the transition to an non-trapped state and leave the cloud. There is an asymmetry to the energy surfaces introduced by the gravitational potential energy. That is, atoms reaching the RF surface near the position, $\vec{r} =[0,0,-z_{\rm{min}}]$ have a lower potential energy than atoms reaching any other point of the RF surface.
\begin{eqnarray}
|z_{\rm{min}}| &=& \frac{h\nu_{\rm{min}}}{h \mu_B g_F |M_F| \left(\frac{dB}{dz}\right)} \nonumber \\
&=& \frac{h\nu_{\rm{min}}}{mg}\frac{I_0}{I}
\end{eqnarray}
Provided that atoms in the trap explore the entire trap volume, the trap depth -- or the maximum energy of the remaining atoms -- is given by,
\begin{eqnarray}
U_{\rm{max}} &=& h\mu_Bg_F|M_F|\left(\frac{dB}{dz}\right)|z_{\rm{min}}|- mg|z_{\rm{min}}| \nonumber \\
&=& h\nu_{\rm{min}}\left[1 - \frac{I_0}{I}\right].
\label{eq:Umax_RF}
\end{eqnarray}

In the present apparatus, the maximum current used to trap the atoms is 200 A, providing a field gradient of 272 G/cm, and the minimum trapping current is $I_0 = 22.4$ A. The maximum RF frequency used here was $\nu_{\rm{max}}=90$~MHz, ejecting atoms with energies up to 3.84 mK, well above the measured trapped ensemble temperature $<$ 1 mK. {The minimum RF frequency used here was $\nu_{\rm{min}}=10$~MHz, which produces a trap depth smaller than 0.1 mK.}

Equation~\ref{eq:Umax_RF} describes the maximum trap depth for a particular minimum RF frequency, assuming that the ensemble of trapped atoms does not have any average energy when loaded into the magnetic trap. In practice, we find that the atoms loaded in the MT have an energy distribution well approximated by a Maxwell-Boltzmann distribution of temperature, T, shifted by an amount $E_{\rm{min}}$.  We believe this shift is due to the offset between the center of the magnetic trap and the MOT. Thus, the actual trap depth is,
\begin{eqnarray}
U = U_{\rm{max}} - \frac{\int^{E_{\rm{max}}}_{E_{\rm{min}}}{E} \  \rho(E-E_{\rm{min}}) dE}{\int^{E_{\rm{max}}}_{E_{\rm{min}}}{\rho(E-E_{\rm{min}}})dE}
\label{eq:Utrap_true}
\end{eqnarray}
where
$\rho(E-E_{\rm{min}})$ is the zero-point shifted Maxwell-Boltzmann distribution describing the trapped ensemble. For each RF frequency, $\nu_{\rm{min}}$, and trap current, $I$, one can calculate the trap depth.  
\subsection{The optimal calibration range for $\ut/\udiff$}

The precision of the determination of $\svtot$ is limited only by the amount of data taken.  However, there is an optimal range of scaled trap depth for this determination. It is clear from Table~\ref{tab:parameters} 
in the main text that the precision of the determination of $\svtot$ improves as the range of $U/\udiff$ for the loss rate measurements increases.  The largest uncertainty is associated with the measurement for Rb-He collisions for which $\utmax/\udiff \simeq 0.07$ and the smallest is for Rb-Xe collisions for which $\utmax/\udiff \simeq 0.4$.  This implies that increasing the trap depth $\utmax$ beyond the maximum of 2.2~mK explored here would improve the precision even further.  However, as $\ut/\udiff$ increases, the retained fraction increases and the systematic uncertainty introduced by sensor ensemble heating increases.  As described earlier, mitigating this effect requires shortening the hold time range which results in a larger uncertainty in the extracted loss rate, $\gloss$.  These two competing effects imply there is an optimal maximum for the scaled trap depth.  Based on the observed variation of the uncertainty in Table~\ref{tab:parameters}
 (which includes fitting uncertainty and the uncertainty due to ensemble heating), the optimal maximum for $\ut/\udiff$ is above 0.25. A quantitative study of the optimum value is a subject for future work.

\section{Quantum Scattering Calculations}

In this section, we provide the details of the quantum scattering calculations of the differential scattering cross sections, the total collision cross sections and $\svl$. 

The scattering event at a given collision energy is described by the $T$-matrix. We compute the $T$-matrices by solving the Schr\"{o}dinger equation using the time-independent coupled channel (CC) approach and the total angular representation of Arthurs and Dalgarno [23]. The method is well described elsewhere [24]. Here, we only provide details pertinent to the calculations in the present work. 

Within the CC approach, the Schr\"{o}dinger equation is reduced to a set of coupled differential equations:
\begin{eqnarray}
\left [ \frac{d^2}{dR^2} - k_{\alpha}^2  + \frac{l(l+1)}{R^2}  \right ] F^J_{\alpha, l; \alpha l} (R) = \sum_{\alpha'} \sum_{l'} U^J_{\alpha, l; \alpha' l'} F^J_{\alpha, l; \alpha' l'} (R),
\label{CC}
\end{eqnarray}
where $R$ is the separation between the centers of mass of the colliding particles, $k_\alpha$ represents the wave number of channel $\alpha$, $l$ is the orbital angular momentum for the rotation of the collision complex,  $J$ is the total angular momentum of the colliding particles and the matrix elements $U^J_{\alpha, l; \alpha', l'}$ are parametrized by the interaction potential of the colliding particles. We integrate these equations by means of the log-derivative [33] and Numerov integration methods.  Eq. (\ref{CC}) are 
solved subject to the scattering boundary conditions and the elements $T_{\alpha l,\alpha'l'}$ of the $T$ matrix are extracted from the asymptotic solutions at large $R$ [23, 24].

For atom - molecule scattering, we treat the molecule as a rigid rotor with rotational angular momentum $j$. In this case, $\alpha = j$. 
The differential scattering cross sections for elastic ($j' = j$) and inelastic ($j' \neq j$) collisions are computed from the $T$-matrix elements as follows: 
\begin{eqnarray}
\frac{d \sigma_{j,j'}}{d \Omega} = \frac{(-1)^{j'-j}}{4(2j+1)k^2_j}\sum_{\lambda = 0}^{\infty} A_{\lambda} P_{\lambda}(\cos \theta), 
\label{differential-glory}
\end{eqnarray}
where $\theta$ is the scattering angle, $P_\lambda$ is a Legendre polynomial of order $\lambda$, and the coefficients $A_{\lambda}$ are given as 
\small
\begin{eqnarray}
A_{\lambda} = \sum_{J_1}^\infty \sum_{J_2}^\infty \sum_{l_1= |J_1 - j|}^{J_1 + j}  \sum_{l_2= |J_2 - j|}^{J_2 + j}  \sum_{l_1'= |J_1 - j'|}^{J_1 + j'}  \sum_{l_2'= |J_2 - j'|}^{J_2 + j'} 
Z(l_1J_1l_2J_2;j\lambda) Z(l_1'J_1l_2'J_2;j'\lambda) T^{J_1 \ast}_{j'l_1';jl_1}  T^{J_2}_{j'l_2';jl_2}, \nonumber \\
\end{eqnarray}
\normalsize
with 
\small
\begin{eqnarray}
Z(abcd;ef) = (-1)^{\frac{1}{2}(f-a+c)} \left [ (2a+1)(2b+1)(2c+1)(2d+1)    \right ]^{1/2} \langle a0,c0|f0\rangle W(abcd;ef),
\end{eqnarray}
\normalsize
where $\langle a0,c0|f0\rangle$ is the Clebsch-Gordan coefficient and $W(abcd;ef)$ is the Racah W-coefficient [34].

The total cross section is computed from the differential cross sections by first integrating over the scattering angle and then summing over all final states of the collision products. 
To calculate the total collision rates, the energy dependence of the total collision cross sections is integrated over the Maxwell-Bolztmann distribution of collision velocities.

The potential energy surface (PES) for atom - rigid rotor interactions is a two dimensional function of $R$ and the Jacobi angle $\chi$ between the vector specifying the direction of the interatomic axis of the molecule and the vector joining the centers of mass of the colliding particles. 
We report calculations with three atom - molecule PESs. Our starting point is a PES that is represented as a Legendre expansion
\begin{eqnarray}
V(R,\chi) = \sum_{s = 0}^6 V_s(R) P_s(\cos \chi). 
\end{eqnarray}
The expansions coefficient $V_{s>0}$ describe the anisotropy of the interaction potential giving rise to inelastic scattering, while the coefficient $V_{s=0}$ is primarily responsible for elastic scattering. Each of the coefficients $V_s$ is represented by the proper  (as permitted by symmetry) long-range expansion
\begin{eqnarray}
V_{s}(R \rightarrow \infty) = \sum_{n} \frac{C_{n,s}}{R^n}
\end{eqnarray}
at large values of $R$. In particular, the isotropic term $V_{s=0}$ is represented at long range as
\begin{eqnarray}
V_{s}(R \rightarrow \infty) = -\frac{C_{6,0}}{R^6} - \frac{C_{8,0}}{R^8} - \frac{C_{10,0}}{R^{10}},
\end{eqnarray}
with $C_{6,0}$ chosen to be $350$ a.u. characteristic of the long-range interaction between Rb atoms and N$_2$ molecules. { The leading anisotropic part of the long-range
interaction is chosen to be $C_{6,2} = 191.04$ a.u.}
These long-range forms are smoothly joined with a short-range repulsive interaction giving the global PES.  The coefficients $V_{s}$ for the starting PES are chosen to generate a global potential that has a minimum of $\approx 235$ cm$^{-1}$ at $R=7.86$ a.u. These parameters are characteristic of van der Waals interactions of closed-shell molecules with alkali metal atoms. We denote this potential surface as PES-I. A cut of this PES is shown by the black line in the inset of FIG. 1(a) of the main text. 
  
The other PESs (hereafter denoted as PES-II and PES-III)  are generated from PES-I by multiplying each of the coefficients $V_s$ by the following function: 
\begin{eqnarray}
f(R) = \frac{ae^{2R} + b}{a e^{2R}}
\end{eqnarray}
with the coefficients $a$ and $b$ chosen such that $f(R) = 1$  when $R > 12$ a.u. for PES-II and when $R>14.2$ a.u. for PES-III. The cross sections of PES-II and PES-III are shown in the inset of FIG. 1(a) of the main text. At $R=5$ a.u., PES-II is magnified by a factor of $102.7$ and PES-III by a factor of $10058.3$. 

For the atom - atom scattering calculations, we approximate the interaction potentials as
\begin{eqnarray}
V(R)  = 4 \epsilon \left[\left(\frac{R_0}{R}\right)^{12} - \left(\frac{R_0}{R}\right)^6\right]\ = \ \left[\frac{C_{12}}{R^{12}} - \frac{C_6}{R^6} \right]
\end{eqnarray}
where the values of the $C_6$ coefficients have been chosen to represent the long-range interactions of the Rb-He, Rb-Ar, Rb-Xe, Li-He, Li-Ar, and Li-Xe systems. The values of the $C_6$ coefficients were borrowed from the literature  [26]. The parameter $C_{12}$ was chosen to ensure a particular value of the energy at the potential energy minimum, as described in the main text.  

For atom - atom scattering calculations, $j=0$ and $J = L$. This reduces Eq.~(\ref{CC}) to a single differential equation and greatly simplifies Eq.~(\ref{differential-glory}) with $j'=j=0$, $J_1 = l_1 = l_1'$ and $J_2 = l_2 = l_2'$ produces the differential scattering cross section $\sigma(v,\theta)$ for given collision velocity $v$ and scattering angle $\theta$. This cross section is used to compute the loss cross section $\sigma_{\rm{loss}}(k, U)$ for each trap depth, $U$. Since the trap loss condition is $U \geq \left(1-\cos\theta\right)\mu^2v^2/\ma$ {(here $\mu=\ma\mb/(\ma+\mb)$ is the reduced mass),} $\sigma_{\rm{loss}}(k, U)$ can be expressed in terms of $\theta$ as 
\begin{eqnarray}
\sigma_{\rm{loss}}(k = \mu v/\hbar, U) &=& \int_{\tmin}^{\pi}\sigma_{\rm{loss}}(v, \theta) \sin\theta \  d\theta .
\label{eq:sigma_intform}
\end{eqnarray}
This cross section is then integrated over the Maxwell-Boltzmann distribution of collision velocities, to yield
\begin{eqnarray}
\svlu & =& \int_0^{\infty}4\pi v^3 \cdot \sigma_{\rm{loss}}(v, U)\cdot \rho(v) \ dv 
\label{eq:svl_MBsum}
\end{eqnarray} 
The rates $\svl$ were computed over trap depths ranging from 0 mK to 15 mK. A data set of $\svl/\svt$ versus $U/U_{\rm{d}}$ was constructed for each collision pair and all six data sets combined and fit to Eq.~(\ref{eq:svl_univ}) by a sixth order polynomial. The results are shown in the first line of Table~2 in the main text of the paper. Over the trap depths investigated in this work, the deviations between the full QS calculations and Eq.~(\ref{eq:svl_univ}) are less than 0.1\%, well below the experimental uncertainty of $\approx$ 1$\%$. Thus the systematic deviations produced by the differences in the interactions between different collision partners are small out to the trap depths investigated, supporting the claim that the expression in Eq. (1) of the main text universally describes the collision loss rate.

\section{Robustness of the Universal Scaling}
\begin{figure}[ht!]
\begin{center}
\includegraphics[scale=0.4]{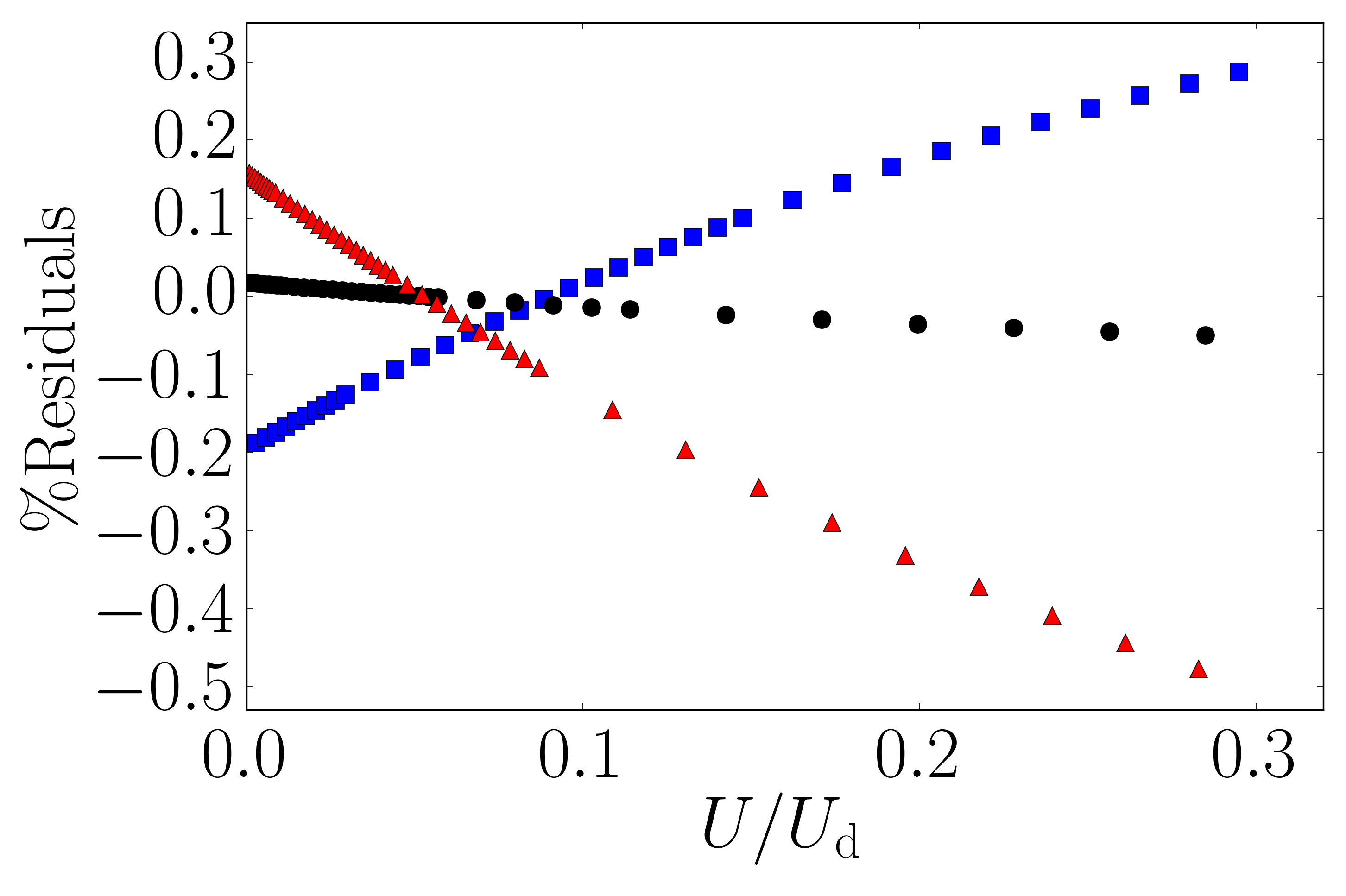}
\caption{A plot the residuals from fitting model $\svl/\svt$ data to the Equation 1 using the best fit coefficients, $\beta_j$, for an ensemble at 294 K as a function of $U/U_d$. The red triangle points correspond to an Ar ensemble at 988 K, the black circles are an Ar ensemble at 294 K, and the blue squares are an Ar ensemble at 3K. The residuals over the range of trap depths accessible to our experiments, $U/U_d \le 0.3$, show a deviation between the model and fit results of less than 0.5\%.  }
\label{fig:TempResiduals}
\end{center}
\end{figure}
The universal curve describing the shape of the loss rate curve as a function of trap depth is robust over a wide range of parameters. The universality arises, in part, from averaging the loss cross-section over collision energy. As a result, the glory oscillations have {little} effect and the loss cross-section depends only on the long range portion of the interaction potential, as described in the main body of the paper. Model loss rate data for Rb-Ar collisions were generated for an Ar ensemble at 3K and a second set for an ensemble temperature of 988K. These data were fit to Equation 1 to find $\svt$ using the $\beta_j$ derived at room temperature. 
The results were remarkable: For the 3K ensemble (blue squares) in Fig.~\ref{fig:TempResiduals}, the actual value for $\svt$ = $7.298\times 10^{-16}$ m$^3$/s was only 0.2\% different from the fitted value of $\svt_{\rm{fit}}$ = $7.312\times 10^{-16}$ m$^3$/s. Similarly, the results for a Ar ensemble at 988K (red triangles) yielded $\svt_{\rm{fit}}$ = $3.916\times 10^{-15}$ m$^3$/s compared to an actual value of $\svt$ = $3.923\times 10^{-15}$ m$^3$/s, again a discrepancy of only 0.2\%. In addition, the residuals of these two, shown in Fig.~\ref{fig:TempResiduals}, still remain below the 0.5 $\%$ level over the entire range of scaled trap depths accessible to our experimental apparatus.

Universality relies on velocity averaging over the glory oscillations which depend on the short range physics. 
The glory oscillations have a period given by $\Delta k \cdot r_0 = 2\pi$ where $r_0 = \left(C_{12}/C_6\right)^{1/6} = \left(C_6/(4\epsilon)\right)^{1/6}$ for a Lennard-Jones potential. Here $\epsilon$ = 50 cm$^{-1}$, is the depth of the potential used in our model, and $r_0$ is the ``range'' of the potential. For our model of Rb-Ar collisions, $r_0 \approx 8 a_0$ and $\Delta k = \hbar/\mu \Delta v$ sets the velocity scale for averaging out the oscillations. For Rb-Ar collisions presented in this section, the glory oscillation period spans a velocity range of $\Delta v \approx 35$ m/s. The FWHM range of the Maxwell-Boltzmann speed distribution is approximately $1.15 \vp$. Since $\bar{v}$ is temperature dependent, we can estimate the lower temperature limit for which the MB distribution FWHM will cover one glory oscillation,
\begin{eqnarray}
\Delta v &\approx& 1.15 \vp \nonumber \\
T_{\rm{min}} &\approx& \frac{m}{2 k_B}\left(\frac{\Delta v}{1.15}\right)^2 \ = \ 2 K.
\label{eq:lowTrange}
\end{eqnarray}

\section{Analytical Modeling of Universal Scaling}
In this section we illustrate analytically why a quasi-universal law for trap loss exists for a pure $-C_6/R^6$ interaction. The collision rate between a sensor atom of mass $\ma$, and {test gas particles of mass} $\mb$, with density, $n$, colliding at a relative speed, $v$, is modeled as,
\begin{eqnarray}
\Gamma &=& n \sigma v.
\label{eq:Gamma_model}
\end{eqnarray}
For elastic collisions where the long-range interaction follows $-C_6/R^6$, the Jeffreys-Born approximation can be used to estimate the angular momentum dependent phase shift and the elastic collision cross-section, 
\begin{eqnarray}
\delta_L &=& \left(\frac{3 \pi}{16} \right)\frac{\mu C_6 k^4}{\hbar^2 L^5} \ = \ \frac{a}{L^5}\\
\sigma ({v}) &=& 8.0828 \left[\frac{C_6}{\hbar {v}} \right]^{\frac{2}{5}} + 7.1889\frac{\hbar}{\mu {v}} \left[\frac{C_6}{\hbar {v}} \right]^{\frac{1}{5}} 
\label{eq:deltaL_C6}
\end{eqnarray}
Here $\mu$ is the reduced mass, and $k = \mu v/\hbar$ is the wavenumber of the reduced mass particle in the center of mass system. Note that the second term in $\sigma({v})$ of Eq.~(\ref{eq:deltaL_C6}) is usually neglected in the literature [25, 32].

Inherently, the gas collision partners impinging on the trapped atoms are characterized by a Maxwell-Boltzmann distribution, $\rho(m, {v, T})$,  at temperature, T, set by the temperature of the walls of the vacuum container. Thus, both the total elastic collision rate, $\Gamma_{\rm{tot}} = n\svt$, and the elastic collision loss rate, $\gloss = n\svlu$, must be averaged over velocity, denoted by $\left<\  \right>$ .  Namely,
\begin{eqnarray}
\svt & = & \left<{v \cdot} \int_{0}^\pi \frac{2\pi}{k^2}\left( \sum_L (2L+1)T_L(k)^*P_L(\cos\theta)\right) \left(\sum_{L^{\prime}} (2L^{\prime}+1)T_{L^{\prime}}(k) P_{L^{\prime}}(\cos\theta) \right) \sin\theta\ d\theta \right> \nonumber \\
 &=& \svt_0 \left[1 + \frac{0.84728}{\vp^{\frac{4}{5}}} \left(\frac{\hbar}{\mu}\right) \left(\frac{\hbar}{C_6}\right)^{\frac{1}{5}}\right] 
\label{eq:svtot_MB}
\end{eqnarray}
where
\begin{eqnarray}
 \svt_0 &=& 8.4946\ \vp^{\frac{3}{5}}\left( \frac{C_6}{\hbar}\right)^{\frac{2}{5}}.
 \end{eqnarray}
 The $T_L(k)$ are the T-matrices associated with the elastic collision process, $\theta$ is the scattering angle of the reduced mass particle in the center of mass frame of the collision, { and $\vp~=~\sqrt{2k_{\rm{B}}T/m}$ is most probable speed for the test gas at temperature, $T$.}
The velocity-averaged loss rate, $\svl$, takes into account the fact that the atoms are held in a trap of depth $U$. That is, in order to be liberated from the trap, the momentum transferred to the trapped atom due to the collision must result in the atom's total energy exceeding the trap depth. In the center of mass frame, this condition reduces to a statement that the reduced mass particle must be scattered outside a minimum angle, $\cos(\tmin) = 1 - \ma U/(\mu^2 {v}^2)$. Thus, $\svl$ is computed in the same manner as $\svt$ from Eq.~(\ref{eq:svtot_MB}) except that the integral ranges from $[\tmin, \pi]$ rather than $[0, \pi]$. Substituting $x = \cos(\theta)$, one has
\begin{eqnarray}
\svl & = & \left<{v \cdot} \int_{-1}^{x_{\rm{min}}} \frac{2\pi}{k^2}\left( \sum_L (2L+1)T_L(k)^*P_L(x)\right) \left(\sum_{L^{\prime}} (2L^{\prime}+1)T_{L^{\prime}}(k) P_{L^{\prime}}(x) \right) dx\right> \nonumber \\
&=& \svt -  \left<{v \cdot} \int_{x_{\rm{min}}}^1\frac{2\pi}{k^2}\left( \sum_L (2L+1)T_L(k)^*P_L(x)\right) \left(\sum_{L^{\prime}} (2L^{\prime}+1)T_{L^{\prime}}(k) P_{L^{\prime}}(x) \right) dx\right> \nonumber \\
\label{eq:svlu_MB}
\end{eqnarray}
For shallow traps, $\tmin$ is less than 10 milliradians which allows the Legendre polynomials in Eq.~(\ref{eq:svlu_MB}) to be expanded, 
\begin{eqnarray}
P_L(x) &\approx& 1 - \frac{L(L+1)}{4}(1-x) + \cdots.
\label{eq:PLexpand}
\end{eqnarray}
{ This approximation for the Legendre polynomials diverges from the small angle approximation used in previous work (for example, see Ref~ \cite{Child}). Both approximations, in addition to the Jeffreys-Born approximation for the scattering-induced phase shift, $\delta_L$, used in $T_L(k)$, limit the partial wave mixing in the analytical equations derived. Therefore, analytical expressions relying on such approximations can only be considered as qualitative indicators of the form of the universal cumulative energy distribution function, $\funiv$.}

The integral over $dx$ in Eq.~(\ref{eq:svlu_MB}) is separated from the summation over partial waves, $L$ and $L^{\prime}$, and from the velocity averaging.  
\begin{eqnarray}
\svlu &\approx& \svt - \left<{\color{blue}v \cdot} \frac{2\pi}{k^2}\sum_{L, L^{\prime}} (2L+1)(2L^{\prime}+1) T_L(k)^* T_L(k)^{\prime} \right. \times \nonumber \\
& & \hspace{30 pt} \left. \left[\left(\frac{\ma U}{\mu^2 v^2}\right)  - \frac{L(L+1) + L^{\prime}(L^{\prime} + 1)}{4} \left( \frac{\ma U}{\mu^2 v^2}\right)^2 + \cdots \right]  \right>
\end{eqnarray}
This description makes it clear that $\svlu$ can be expanded in powers of $U$ for shallow traps. {The same qualitative conclusion can be drawn starting with previous estimates of the shallow angle scattering amplitudes \cite{Child}.}

The exact {analytical } form of the expansion will depend on the T-matrix, $T_L(k)$, which encodes the {nature} of the long range interaction into the velocity averaged loss rate. Using the Jeffreys-Born approximation, 
\begin{eqnarray}
T_L(k) &= & \sin(\delta_L) \cos(\delta_L) + i \sin^2(\delta_L) \nonumber \\
 &=& \frac{1}{2}\sin\left(\frac{2a}{L^5} \right) + i \sin^2\left(\frac{a}{L^5} \right)
 \label{eq:Tlk_JB}
 \end{eqnarray}
 where the phase shift is given by Eq.~(\ref{eq:deltaL_C6}). This form of the phase shift is only valid when the velocity-dependent phase associated with core repulsion scattering, leading to glory oscillations, are eliminated through velocity averaging. 
 
Performing the integrations with this phase shift leads to,
\begin{eqnarray}
\svl &\approx & \svt - \alpha_1 \left( \frac{\ma U}{\hbar^2}\right) + \alpha_2 \left( \frac{\ma U}{\hbar^2}\right)^2 + \cdots
\label{eq:svl_alphas}
\end{eqnarray}
where,
\begin{eqnarray}
\alpha_1 &=& \frac{0.0554929\svt^2}{\vp}\left[1 - \epsilon_1 \right] \ = \ \gamma_1   \frac{\svt^2}{\vp}\\
	\alpha_2 &=& \frac{{0.004315}\svt^3}{\vp^2}\left[ 1 - \epsilon_2 \right] \ = \ \gamma_2  \frac{\svt^3}{\vp^2}
\label{eq:alphas}
\end{eqnarray}
Here,
\begin{eqnarray}
\epsilon_1 &=& \left(\frac{\svt_0}{\svt}\right)^2 \left[ \frac{0.2158}{\vp^{\frac{4}{5}}} \left(\frac{\hbar}{\mu}\right)\left(\frac{\hbar}{C_6} \right)^{\frac{1}{5}} - 
\frac{0.02795}{\vp^{\frac{8}{5}}} \left(\frac{\hbar}{\mu}\right)^2\left(\frac{\hbar}{C_6} \right)^{\frac{2}{5}} \right]
\label{eq:epsilon1}
\end{eqnarray}
and 
\small
\begin{eqnarray}
\epsilon_2 &=& \left(\frac{\svt_0}{\svt}\right)^3 \left[ \frac{{0.6827}}{\vp^{\frac{4}{5}}} \left(\frac{\hbar}{\mu}\right)\left(\frac{\hbar}{C_6} \right)^{\frac{1}{5}} + 
\frac{{0.3356}}{\vp^{\frac{8}{5}}} \left(\frac{\hbar}{\mu}\right)^2\left(\frac{\hbar}{C_6} \right)^{\frac{2}{5}} -
\frac{{0.7891}}{\vp^{\frac{12}{5}}} \left(\frac{\hbar}{\mu}\right)^3 \left(\frac{\hbar}{C_6} \right)^{\frac{3}{5}}
\right] \nonumber \\
\label{eq:epsilon2}
\end{eqnarray}
\normalsize
Combining these equations illustrates the emergence of the quasi-universal behaviour,
\begin{eqnarray}
\svl &=& \svt \left[1 - 4\pi \gamma_1 \left(\frac{\ma \svt/\vp}{4\pi \hbar^2 } \right)U  + (4\pi)^2 \gamma_2 \left(\frac{\ma \svt/\vp}{4\pi \hbar^2 } \right)^2 U^2 + \cdots\right] \nonumber \\
&=& \svt \left[1 - 4\pi \gamma_1 \left(\uud\right)  + (4\pi)^2 \gamma_2 \left(\uud\right)^2 + \cdots \right] \nonumber \\
&=& \svt \left[1 - \beta_1 \left(\uud\right) + \beta_2\left(\uud\right)^2 - \cdots \right] \ = \ \svt\left(1-\funiv \right)
\label{eq:svl_univ}
\end{eqnarray}
In Eq.~(\ref{eq:svl_univ}), the quantum diffractive energy has been defined as \cite{PhysRevA.60.R29},
\begin{eqnarray}
U_{\rm{d}} &=& \frac{4\pi\hbar^2}{\ma \svt/\vp}.
\label{eq:Ud}
\end{eqnarray}

The universality of the coefficients in Eq.~(\ref{eq:svl_univ}) is disrupted by the $\epsilon_i$ terms, defined in Eqs.~(\ref{eq:epsilon1}) and (\ref{eq:epsilon2}) for $i = 1, 2$. These terms introduce a dependence on the room-temperature collision partner through the most probable velocity, $1/\vp^{n/5}$, terms. For lighter collision partners, these become more significant. Further, there are $(1/C_6)^{n/5} \times (1/\mu)^n$ terms which introduce some dependence on the trapped atom mass and the long-range van der Waals coefficient. For lower reduced masses and smaller $C_6$ coefficients, these corrections are more significant.

Table~\ref{tab:bjanalytical} lists the values of $\beta_1$ and $\beta_2$ determined from the full quantum scattering computations, derived from the analytical expression for the small angle, elastic scattering amplitude reported in Ref.~\cite{Child}, and from the analytical predictions provided here.
\begin{table}[ht!]
\centering
\begin{tabular}{|c|c|c|}
\hline
\hline
 & $\beta_1$ & $\beta_2$\\
\hline
\hline
Full Numerical Computation & 0.6730(7) & -0.477(3) \\ 
\hline
\hline
Values derived from Ref. \cite{Child} & 0.764 & -0.791 \\
\hline
\hline
This Appendix &\ \  0.693(3)\ \  &\ \  -0.669(9)\ \  \\
\hline
\hline
\end{tabular}
\caption{The valued of $\beta_1$ and $\beta_2$ from the full quantum scattering computations (averaged over Rb-[He,Ar,Xe] and Li-[He, Ar, Xe] collisions), derived from the analytical expression for the small angle, elastic scattering amplitude from Ref.~\cite{Child}, and  from the analytical expressions used in this Appendix (averaged over Rb-[He, Ar, Xe] and Li-[He, Ar, Xe] collisions).
}
\label{tab:bjanalytical}
\end{table}

\subsection{Generalization to $V(R) = -C^n/R^n$ Long-Range Potentials}

The above methods can be generalized to other forms of long range potential, in particular to $V(R) = -C_n/R^n$ for n = 3, 4, 5, etc.  To begin, one generalizes the approximate angular momentum dependent phase shift,
\begin{eqnarray}
\delta_L^{(n)} & = & \left(\frac{\mu\ C_n\ k^{n-2}}{\hbar^2 \ L^{n-1}}\right)  \left( \frac{\sqrt{\pi}\ \Gamma(\frac{n-1}{2})}{2 \Gamma(\frac{n}{2})}\right) \nonumber \\
&=& \left(\frac{\mu\ C_n\ k^{n-2}}{\hbar^2 \ L^{n-1}}\right) \xi(n) 
\label{eq:delta_L_n}
\end{eqnarray}
Table~\ref{tab:zivals} provides the values for the $\xi(n)$ function for various values of $n$. 

The corresponding cross-sections are,
\begin{eqnarray}
\sigma(k,n) &=& \frac{2\pi}{k^2}\left[\left(2 a(n)\right)^{\frac{2}{n-1}} \cos\left(\frac{\pi}{n-1}\right) \Gamma\left( \frac{n-3}{n-1}\right)  \right. \nonumber \\ 
 & & \hspace{12pt} \left. + \left(2 a(n) \right)^{\frac{1}{n-1}} \cos\left(\frac{\pi}{2(n-1)}\right) \Gamma\left( \frac{n-2}{n-1}\right) \right] \nonumber \\
 &=& 2\pi \left[ \left(2 \xi(n) \frac{C_n}{\hbar v}\right)^{\frac{2}{n-1}} \cos\left(\frac{\pi}{n-1}\right) \Gamma\left(\frac{n-3}{n-1}\right) \right.\nonumber \\
 & & \hspace{12pt} \left. + \   \left(\frac{\hbar}{\mu v}\right) \left(2 \xi(n) \frac{C_n}{\hbar v}\right)^{\frac{1}{n-1}} \cos\left(\frac{\pi}{2(n-1)}\right) \Gamma\left(\frac{n-2}{n-1}\right)\right] \nonumber \\
 &=& c(n) \left(\frac{C_n}{\hbar v}\right)^{\frac{2}{n-1}} + d(n) \left(\frac{\hbar}{\mu v}\right) \left(\frac{C_n}{\hbar v}\right)^{\frac{1}{n-1}}
 \label{eq:genearal_sigma_v}
\end{eqnarray}
which are valid for $n>3$.  The expressions are listed below in Table~\ref{tab:zivals}.

The velocity averaged total elastic collision cross-sections are, 
\begin{eqnarray}
\svtn &=& \frac{2}{\sqrt{\pi}}\left[ c(n) \vp^{\frac{n-3}{n-1}} \Gamma\left(\frac{2n-3}{n-1}\right) \left(\frac{C_n}{\hbar}\right)^{\frac{2}{n-1}} + d(n) \frac{1}{\vp^{\frac{1}{n-1}}} \left(\frac{\hbar}{\mu}\right) \Gamma\left(\frac{3n-4}{2(n-1)}\right) \left(\frac{C_n}{\hbar}\right)^{\frac{1}{n-1}} \right] \nonumber \\
 &=& \svtn_0 \left[1+\frac{d(n)}{c(n)} \frac{1}{\vp^{\frac{n-2}{n-1}}}\frac{\Gamma\left(\frac{3n-4}{2(n-1)}\right)}{\Gamma\left(\frac{2n-3}{n-1}\right)} \left(\frac{\hbar}{\mu}\right) \left(\frac{\hbar}{C_n}\right)^{\frac{1}{n-1}} \right]
\label{eq:svtn}
\end{eqnarray}
with the expressions summarized in column 3 of Table~\ref{tab:zivals}. One observes that for all forms $V(R) = -C_n/R^n$, the description for $\svtn$ follows the format described above for $n =6$. Similarly, the description for $\svl$ provided in Equation~\ref{eq:svl_univ} also applies to these long range potentials, with unique values for the expansion coefficients, $\beta_j^{(n)}$.

\begin{table}[ht!]
\centering
\begin{tabular}{|c|c|c|c|c|c|c|}
\hline
n & 3 &4 & 5 & 6 & 8 & 10 \\
\hline
$\xi(n)$ & 1 &$\frac{\pi}{4}$ &$\frac{2}{3}$ &$\frac{3\pi}{16}$ &$\frac{5\pi}{32}$ &$\frac{35\pi}{256}$ \\
\hline
$c(n)$&-&10.0823&8.8352&8.0828&7.1703&6.6126\\
\hline
$d(n)$&-&8.0648&7.5347&7.1889&6.7486&6.4693\\
\hline
\end{tabular}
\caption{The values of $\xi(n)$, $c(n)$, and $d(n)$ as a function of $n$ for potentials of the form, $V(R) = - C_n/R^n$.
}
\label{tab:zivals}
\end{table}

\clearpage


\end{document}